\let\ifarxiv=\iffalse    
\pdfoutput=1

\ifarxiv

\documentclass[12pt,a4paper]{article}
\usepackage[a4paper,text={450pt,650pt},centering]{geometry}

\fi

\ifarxiv\else

\documentclass[11pt,a4paper]{article}
\usepackage{mathptmx}
\usepackage[a4paper,text={130mm,198mm}]{geometry}

\fi

\usepackage{amsmath,amssymb}
\usepackage[bookmarks=true,hyperfigures=true]{hyperref}
\usepackage{graphicx}
\usepackage[nosort]{cite}
\ifarxiv\usepackage[bulletsep]{collref}\fi
\usepackage{multirow}
\usepackage{amsfonts}
\usepackage{amssymb}
\usepackage{amscd}

\let\oldbfseries=\bfseries
\let\oldmdseries=\mdseries
\let\oldnormalfont=\normalfont
\renewcommand{\bfseries}{\oldbfseries\boldmath}
\renewcommand{\mdseries}{\oldmdseries\unboldmath}
\renewcommand{\normalfont}{\oldnormalfont\unboldmath}

\allowdisplaybreaks[3]

\numberwithin{equation}{section}

\usepackage[font=small,labelfont=bf,width=0.85\textwidth]{caption}

\providecommand{\hypersetup}[1]{}
\providecommand{\texorpdfstring}[2]{#1}

\hypersetup{plainpages=false}
\hypersetup{pdfpagemode=UseNone}
\hypersetup{bookmarksnumbered=true}
\hypersetup{pdfstartview=FitH}
\hypersetup{colorlinks=false}
\hypersetup{citebordercolor={.5 1 .5}}
\hypersetup{urlbordercolor={.5 1 1}}
\hypersetup{linkbordercolor={1 .7 .7}}

\providecommand{\arxivref}[2]{\href{http://arxiv.org/abs/#1}{#2}}

\providecommand{\href}[2]{#2}
\providecommand{\arxivlink}[1]{\href{http://arxiv.org/abs/#1}{arxiv:#1}}

\begin{document}


\thispagestyle{empty}
\phantomsection
\addcontentsline{toc}{section}{Title}

\begin{flushright}\footnotesize%
\texttt{\arxivlink{1012.4003}}\\
overview article: \texttt{\arxivlink{1012.3982}}%
\vspace{1em}%
\end{flushright}

\begingroup\parindent0pt
\begingroup\bfseries\ifarxiv\Large\else\LARGE\fi
\hypersetup{pdftitle={Review of AdS/CFT Integrability, Chapter V.3: Scattering Amplitudes at Strong Coupling}}%
Review of AdS/CFT Integrability, Chapter V.3:\\
 Scattering Amplitudes at Strong Coupling
\par\endgroup
\vspace{1.5em}
\begingroup\ifarxiv\scshape\else\large\fi%
\hypersetup{pdfauthor={Luis F. Alday}}%
Luis F.\ Alday
\par\endgroup
\vspace{1em}
\begingroup\itshape
Mathematical Institute, University of Oxford\\
24--29 St.\ Giles', Oxford OX1 3LB, UK.\par\vspace{1em}
School of Natural Sciences, Institute for Advanced Study,\\
Princeton, NJ 08540, USA
\par\endgroup
\vspace{1em}
\begingroup\ttfamily
alday@maths.ox.ac.uk
\par\endgroup
\vspace{1.0em}
\endgroup

\begin{center}
\includegraphics[width=5cm]{TitleV3.mps}
\vspace{1.0em}
\end{center}

\paragraph{Abstract:}
We review the computation of scattering amplitudes of planar
maximally super-symmetric Yang-Mills at strong coupling. By
using the $AdS/CFT$ duality the problem boils down to the
computation of the area of certain minimal surfaces on $AdS$.
The integrability of the model can then be efficiently used in
order to give an answer for the problem in terms of a set of
integral equations.

\ifarxiv\else
\fi
\hypersetup{pdfsubject={MSC (2010): 81T60, 81T30, 49Q05}}%

\ifarxiv\else
\fi
\hypersetup{pdfkeywords={Scattering Amplitudes, AdS/CFT duality, Integrability.}}%

\newpage


\section{Introduction} \label{sec:intro}

The aim of this review is to study gluon scattering amplitudes of four dimensional planar maximally super-symmetric Yang-Mills (MSYM).  We hope that the study of such amplitudes would teach us something about scattering amplitudes of QCD, but at the same time they are much more tractable. The reason for such tractability is twofold. On one hand, perturbative computations are much simpler than in QCD, due to the high degree of symmetry. In fact enormous progress has been made in the last few years. On the other hand, the strong coupling regime of the theory can be studied by means of the AdS/CFT duality, by studying a weakly coupled string sigma-model.

In this review we focus on how to use the AdS/CFT duality in order to compute gluon scattering amplitudes of planar MSYM at strong coupling, referring the reader to \cite{chapAmp,chapDual} for details on the perturbative side of the computation . In section two we set up the problem of computing scattering amplitudes at strong coupling. The problem boils down to the computation of the area of certain minimal surfaces in $AdS$. For the particular  case of four gluons, such surface, and its area, can be explicitly computed. Furthermore, the strong coupling computation hints at some symmetries that actually appear to be symmetries at all values of the coupling. This is briefly reviewed at the end of section two. In section three we focus on the mathematical problem of computing the area of minimal surfaces in $AdS$. The integrability of the model allows the introduction of a spectral parameter. By studying the problem as a function of the spectral parameter we are able to give a solution in the form of a set of integral equations. These equations have the precise form of thermodynamic Bethe ansatz (TBA) equations. The area turns out to coincide with the free energy of such TBA system. Finally, In section four, we end up with some conclusions and a list of open problems.

\section{Gluon scattering amplitudes at strong coupling}

Four dimensional MSYM, the theory whose amplitudes we want to
consider, turns out to be dual to type IIB string theory on
$AdS_5 \times S^5$. This duality receives the name of
$AdS/CFT$ duality \cite{Maldacena:1997re} and is the main focus of this review.
%
%
A remarkable feature of this duality is that it allows
to compute certain observables of MSYM at strong coupling by
doing geometrical computations on $AdS$. A well known example
is the computation of the expectation value of super-symmetric
Wilson loops, which reduces to a minimal area problem
\cite{Rey:1998ik,Maldacena:1998im}. In this section we will
show that this is also the case for the computation of
scattering amplitudes at strong coupling! \footnote{In this
section, we follow closely \cite{Alday:2007hr}, to which we
refer the reader for the details.}

As in the gauge theory, we will need to introduce a regulator
in order to define properly scattering amplitudes. In order to
set-up our computation we introduce a D-brane as IR regulator,
as we explain in detail below. Another convenient regulator is
the strong coupling/super-gravity analog of dimensional
regularization. This regulator will be used in order to compare our results with expectations from the perturbative side.

\subsection{Set-up of the computation}

In order to set up the computation at strong coupling, it is convenient to introduce a regularization as follows. We start from a $U(N+k)$ theory, with $k \ll N$, and then consider a vacuum breaking the symmetry to $U(N) \times U(k)$ by giving to a scalar field a vacuum expectation value $m_{IR}$ which plays the role of an infrared cut-off.\footnote{See \cite{Alday:2009zm} for perturbative computations using this regulator.} When we take the 't Hooft limit we keep $k$ fixed, so that the low energy $U(k)$ theory becomes free. We then scatter gluons of this $U(k)$ theory. We are interested in the regime where all kinematic invariants are much larger than the IR cut-off, $s_{ij} \gg m_{IR}^2$. It turns out that the leading exponential behavior at strong coupling can be captured simply by considering $k=1$. At strong coupling this corresponds to consider a $D3$-brane localized in the
radial direction. More precisely, we start with the $AdS_5$
metric written in Poincare coordinates

\begin{equation}
\label{origads}
ds^2 = R^2 {dx_{3+1}^2 +dz^2 \over z^2}
\end{equation}
and place a $D3$-brane at some fixed large value of $z=z_{IR} $
and extending along the $x_{3+1}$ coordinates. The asymptotic
states are open strings that end on that D-brane. We then
consider the scattering of these open strings, that will have
the interpretation of the gluons that we are scattering.

The proper momentum of the strings is $k_{pr}=k z_{IR}/R$, where
$k$ is the momentum conjugate to $x_{3+1}$, plays the role of
gauge theory momentum and will be kept fixed as we take away the
IR cut-off, $z_{IR} \to \infty$. Therefore, due to the warping of the metric, the
proper momentum is very large, so we are considering the
scattering of strings at fixed angle and with very large momentum.

Amplitudes in such regime were studied in flat space by Gross and
Mende \cite{Gross:1987ar}. The key feature of their computation is that the amplitude
is dominated by a saddle point of the classical action. In our
case we need to consider classical strings on $AdS$. Hence, we need to consider a world-sheet with the topology of a disk
with vertex operator insertions on its boundary, which correspond
to the external states (see fig. 1).  A disk amplitude with a fixed ordering of the open string vertex operators corresponds to a given color ordered amplitude.

\begin{figure}[h]
\centering
\includegraphics[scale=0.35]{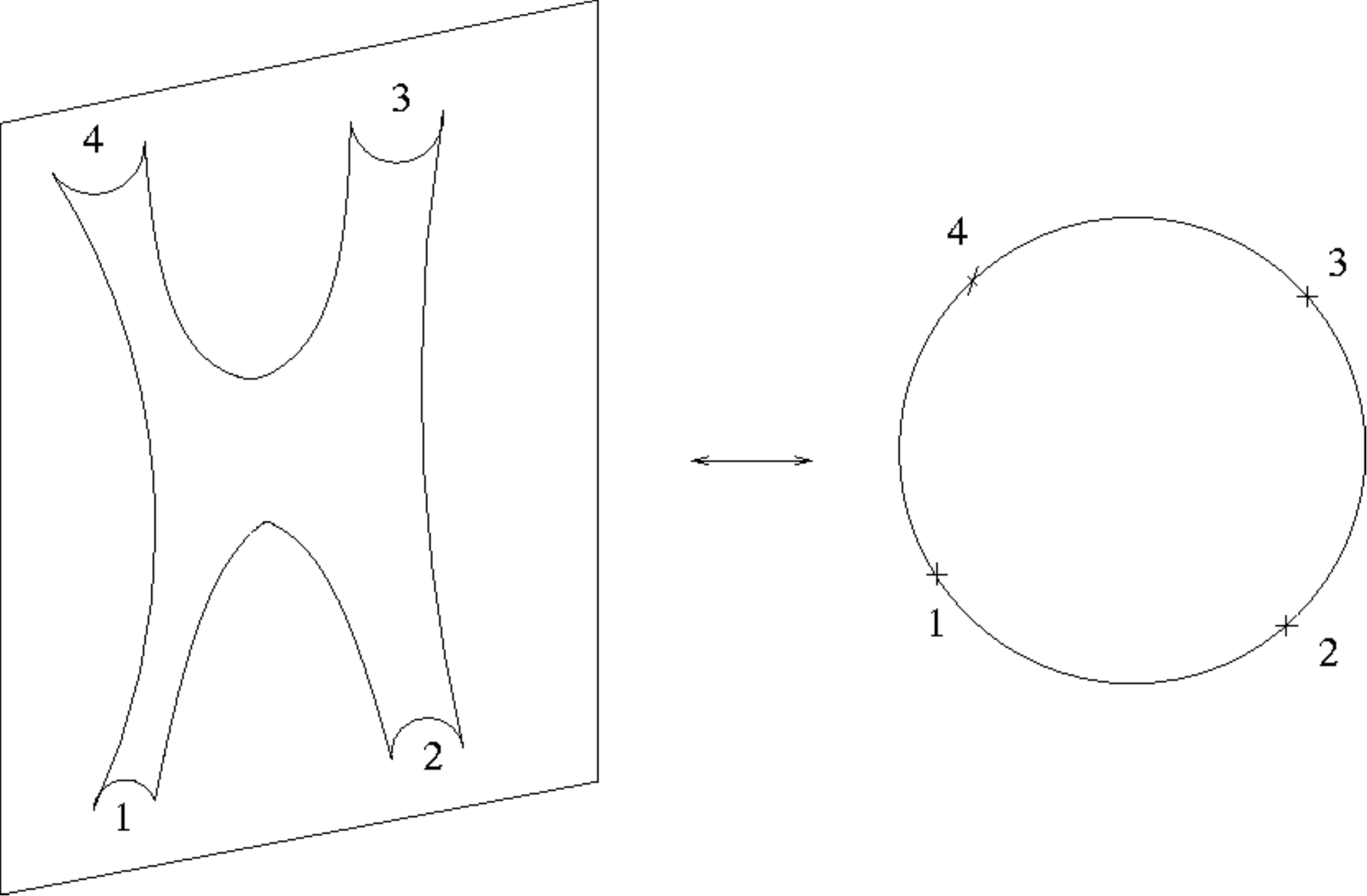}
\caption{World-sheet corresponding to the scattering of four open
strings. In the figure on the left we see four open strings ending on the IR D-brane, the world-sheet has then the topology of a disk, shown on the right, with four vertex operator insertions.} 
\label{fig:1}
\end{figure}

What are the boundary conditions for such world-sheet? since the open strings are attached to the D-brane, $z=z_{IR}$
at the boundary. Furthermore, in the vicinity of a vertex operator, the momentum of the
external state should fix the form of the solution.

In order to state more simply the boundary conditions for the
world-sheet, it is convenient to describe the solution in terms
of T-dual coordinates $y^\mu$, defined as follows

\begin{equation}
\label{tdual}
ds^2=w^2(z)dx_\mu dx^\mu +... ~~~\rightarrow ~~~
\partial_\alpha y^\mu = i w^2(z) \epsilon_{\alpha \beta} \partial_\beta x^\mu
\end{equation}
The presence of the $i$ is due to the fact that we are considering a Euclidean world-sheet in Minkowski space-time. Note that we do not T-dualize along the radial direction. After
defining $r=R^2/z$ the dual metric takes the form

\begin{equation}
\label{dualads}
ds^2=R^2 {dy_\mu dy^\mu+dr^2 \over r^2}
\end{equation}
Note that this metric is equivalent to the same $AdS_5$ metric
we started with! A crucial difference is that now, in terms of
the dual coordinates, the boundary of the
world-sheet is located at $r=R^2/z_{IR}$, which is very small. Furthermore, the $T-$duality we performed interchanges Neumann by Dirichlet boundary conditions. This means that the boundary of the world-sheet sits at a fixed point in the space of the dual coordinates. When a vertex operator with momentum $k^\mu$ is inserted, the location of such point gets shifted by an amount proportional to $\Delta y^\mu=2 \pi k^\mu$.

Summarizing, the boundary of the world-sheet is located at
$r=R^2/z_{IR}$ and is a particular line constructed as follows

\begin{itemize}
\item For each particle of momentum $k^\mu$, draw a
segment joining two points separated by $\Delta y^\mu=2 \pi
k^\mu$.
\item Concatenate the segments according to the insertions
    on the disk.
\end{itemize}

Since gluons are massless, the segments are light-like.
Furthermore, due to momentum conservation, the segments form a
closed polygon. The world-sheet, when expressed in T-dual
coordinates, will then end in such sequence of light-like
segments (see fig. 2) located at $r=R^2/z_{IR}$.

\begin{figure}[h]
\centering
\includegraphics[scale=0.3]{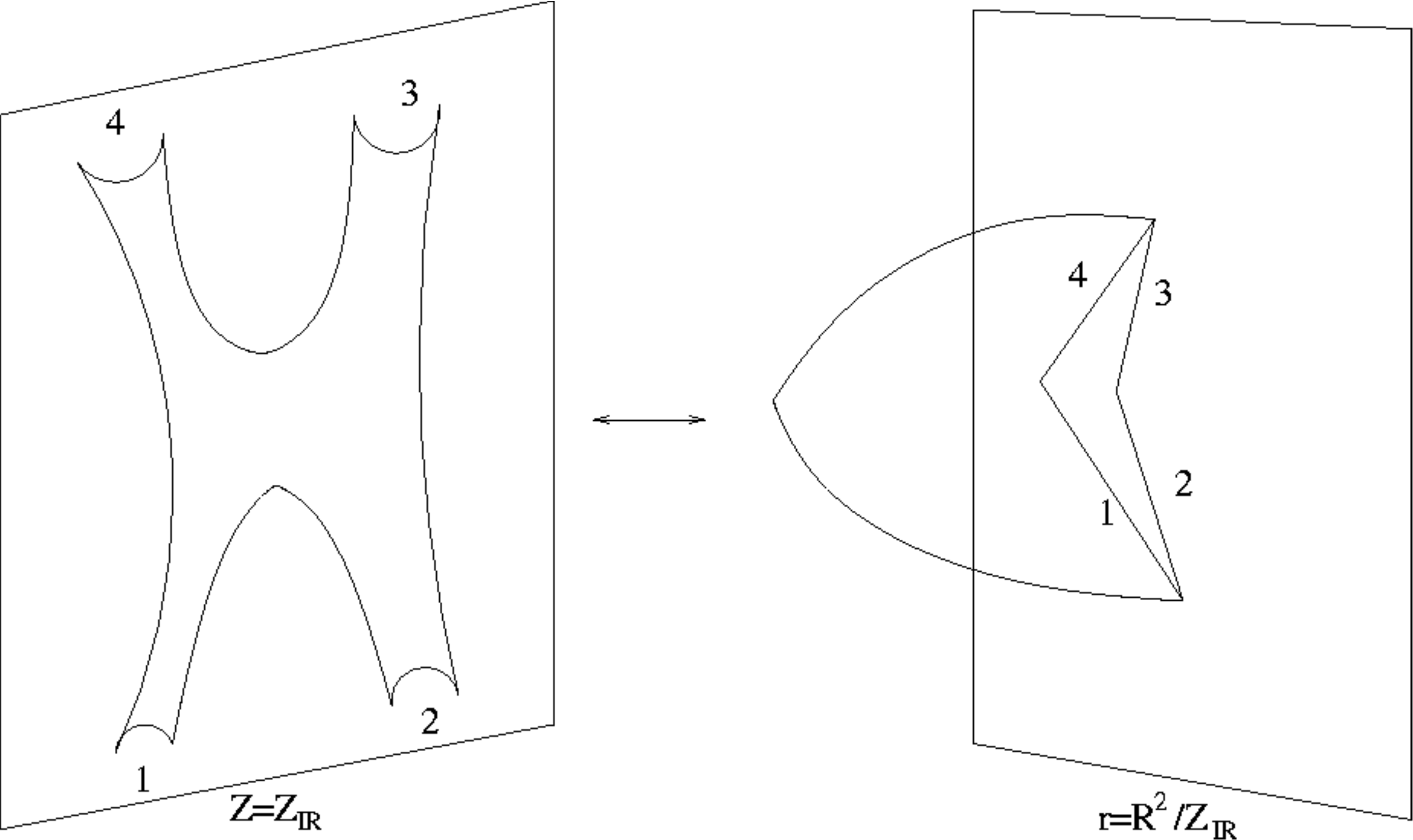}
\caption{Comparison of the world sheet in original and
T-dual coordinates. The hyperplane on the picture to the right should not be interpreted as a D-brane, but rather as a radial slice where the boundary of the world-sheet it located.} 
\label{fig:2}
\end{figure}

As we take away the IR cut-off, $z_{IR} \rightarrow \infty$,
the boundary of the world-sheet moves towards the boundary of
the T-dual metric, at $r=0$. This computation would then be formally equivalent to the computation of the
expectation value of a Wilson loop given by a
sequence of light-like segments at strong coupling \cite{Rey:1998ik,{Maldacena:1998im}}. \footnote{As explained in detail in \cite{chapDual}, this remarkable duality between Wilson loops and scattering amplitudes was also observed in perturbative computations.}

Our prescription is that the leading exponential behavior of
the $N-$point scattering amplitude is given by the area $A$ of the
minimal surface that ends on a sequence of light-like segments on
the boundary

\begin{equation}
\label{finalpresc}
{\cal A}_N \sim e^{-{\sqrt{\lambda} \over 2
\pi}A(k_1,...,k_N)}
\end{equation}
An important comment is in order. Note that the strong coupling
computation is blind to the type or polarization of the
external particles. Such information will contribute to
prefactors in (\ref{finalpresc}) and will be subleading in a
$1/\sqrt{\lambda}$ expansion, relative to the leading exponential term\footnote{As discussed in more detail in \cite{McGreevy:2008zy}, when computing the disk amplitude in the saddle point approximation, one can neglect the polarization of the gluon vertex operators.}.
These differences should be
visible once we consider quantum corrections to the classical
area. This is still an open problem.

Some generalizations to the above picture were developed. In \cite{finitetemp,finitetemp2} finite temperature was introduced while in \cite{Dorn:2009hs} the authors considered solutions with non trivial motion on the $S^5$. Finally, in \cite{quarks,quarks2,quarks3}, the scattering of quarks at strong coupling was considered. Unfortunately, due to space limitations, we wont discuss these interesting developments here, but refer the reader to the original literature.

We have then reduced the problem of computing scattering
amplitudes at strong coupling to the problem of finding minimal
surfaces in $AdS$. In the following we will show that such
surface can be found for the particular case of the scattering
of four gluons. To find and understand this solution in detail
will be quite instructive. Then, in the next section, we will
use the integrability of the problem in order to give a general
solution, for any number of gluons, in the form of a set of
integral equations.

\subsection{Scattering of four gluons}

Consider the scattering of two particles into two particles,
$k_1+k_3 \rightarrow k_2+k_4$ and define the usual Mandelstam
variables

\begin{equation} s=-(k_1+k_2)^2,~~~~~t=-(k_2+k_3)^2\end{equation}

According to our prescription we need to find the minimal surface
ending in the following light-like polygon

\begin{figure}[h]
\centering
\includegraphics[scale=0.4]{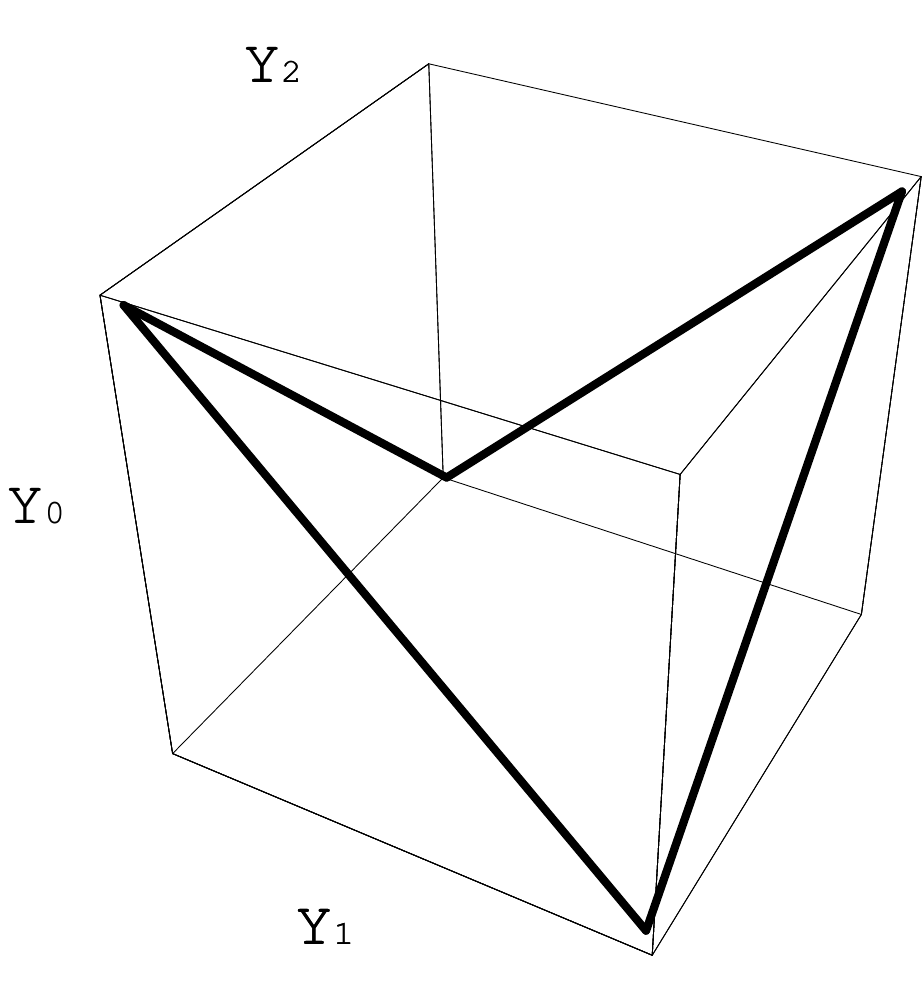}
\caption{Polygon corresponding to the scattering of four gluons}
\end{figure}

In order to write the Nambu-Goto action it is convenient to use
Poincare coordinates $(r,y_0,y_1,y_2)$, setting $y_3=0$ and
parametrize the surface by its projection to the $(y_1,y_2)$
plane. In this case we obtain an action for two fields, $r$ and
$t$, living in the space parametrized by $y_1$ and $y_2$
\begin{equation} S = {R^2 \over 2 \pi } \int dy_1 dy_2 { \sqrt{ 1 + (
\partial_i r)^2 - (
\partial_i y_0)^2 - (
\partial_1 r \partial_2y_0 - \partial_2 r \partial_1 y_0 )^2 } \over
r^{ 2 } } \end{equation}
where $r=r(y_1,y_2)$ and $\partial_i r=\partial_{y_i} r$, etc.
Our aim is to find a solution to the classical equations of
motion with the appropriate boundary conditions. Let us
consider first the case $s=t$, where the projection of the
polygon lines to the $(y_1,y_2)$ plane
 is a square. By scale invariance we can choose the edges of the
square to lie at $y_1,y_2=\pm 1$. The boundary conditions are
then given by

\begin{equation}
\label{bcsquare}
 r(\pm 1,y_2)=r(y_1,\pm 1)=0,~~~~y_0(\pm 1,y_2)=\pm
y_2,~~~y_0(y_1,\pm 1)=\pm y_1 \end{equation}
In \cite{Kruczenski:2002fb} the solution corresponding to a
single cusp was considered. One can make educated guesses using
such solution as a guidance and propose
\begin{equation} \label{squaresol} y_0(y_1,y_2)=y_1
y_2,~~~~~r(y_1,y_2)=\sqrt{(1-y_1^2)(1-y_2^2)}\end{equation}
Remarkably this turns out to be a solution of the equations of
motion! This is the solution for the case $s=t$, how can we obtain the most general solution?

The dual $AdS_5$ space has a $SO(2,4)$ group of isometries. This symmetry is sometimes referred to as "dual conformal
symmetry" and should not be confused with the
original $SO(2,4)$ symmetry associated to the original $AdS$ space. This dual symmetry can be used in order to map the particular solution we have just found to the most general solution with four edges, in particular with $s \neq t$.
The general solution can be conveniently written as
\begin{eqnarray}r={a \over \cosh u_1 \cosh u_2+b \sinh u_1 \sinh u_2},~~~~
y_0= {a \sqrt{1+b^2} \sinh u_1 \sinh u_2 \over \cosh u_1 \cosh
u_2+b \sinh u_1 \sinh u_2} \\ y_1={a \sinh u_1 \cosh u_2 \over
\cosh u_1 \cosh u_2+b \sinh u_1 \sinh u_2},~~~~ y_2={a \cosh u_1
\sinh u_2 \over \cosh u_1 \cosh u_2+b \sinh u_1 \sinh u_2}
\end{eqnarray}
where $u_{1,2}$ parametrize the world-sheet and we have written the surface as a solution to the equations
of motion in conformal gauge.
%
$a$ and $b$ encode the kinematical information of the scattering
as follows

\begin{equation}-s(2 \pi)^2 = {8 a^2 \over (1-b)^2},~~~~~-t (2 \pi)^2 ={8 a^2
\over (1+b)^2},
 ~~~~~{ s \over t } = { (1+b)^2 \over
(1-b)^2 } \end{equation}

According to the prescription, we should now plug the classical
solution into the classical action to compute the area and obtain the four
point scattering amplitude at strong coupling. However, in
doing so, we obtain a divergent answer. That is of course the
case, since we have taken the IR regulator away. In order to
obtain a finite answer we need to reintroduce a regulator.
Since we want to compare our results to field theory
expectations, it is convenient to introduce the strong
coupling analog of dimensional regularization.

Gauge theory amplitudes are regularized by considering the
theory in $D=4-2\epsilon$ dimensions. More precisely, one
starts with ${\cal N}=1$ in ten dimensions and then
dimensionally reduce to $4-2\epsilon$ dimensions. For integer
$2\epsilon$ this is precisely the low energy theory living on a
$Dp-$brane, where $p=3-2\epsilon$. We regularize the amplitudes
at strong coupling by considering the gravity dual of these
theories and then analytically continuing in $\epsilon$. The
string frame metric is
 \begin{eqnarray}ds^2=f^{-1/2}dx_{4-2\epsilon}^2+f^{1/2}\left[ dr^2+r^2 d\Omega^2_{5+2\epsilon} \right],~~~~~f=(4 \pi^2 e^\gamma)^\epsilon \Gamma(2+\epsilon) \mu^{2\epsilon}
 \frac{\lambda}{r^{4+2\epsilon}}\end{eqnarray}
Following the steps described above, we are led to the following
action \begin{equation}S=\frac{\sqrt{c_\epsilon \lambda}
\mu^\epsilon }{2\pi} \int \frac{{\cal
L}_{\epsilon=0}}{r^\epsilon}\end{equation}

Where ${\cal L}_{\epsilon=0}$ is the Lagrangian density for
$AdS_5$. The presence of the factor $r^\epsilon$ will have two
important effects. On one hand, previously divergent integrals
will now converge (if $\epsilon < 0$). On the other hand, the
equations of motion will now depend on $\epsilon$ and we were
not able to compute the full solution for arbitrary $\epsilon$.
However, we are interested in computing the amplitude only up
to finite terms as we take $\epsilon \rightarrow 0$. In that
case, it turns out to be sufficient to plug the original
solution into the $\epsilon$-deformed action \footnote{Up to a
contribution from the regions close to the cusps that adds an
unimportant additional constant term.}. After performing the
integrals and expanding in powers of $\epsilon$ we get the
final answer

\begin{eqnarray} \label{finalfour}{\cal A}=e^{-\frac{\sqrt{\lambda}}{2\pi}A},~~~~,-\frac{\sqrt{\lambda}}{2\pi}A=i
S_{div}+\frac{\sqrt{\lambda}}{8\pi}\left(\log{\frac{s}{t}}
\right)^2+\tilde{C} \\ \nonumber S_{div}=2S_{div,s}+2S_{div,t}\\ \nonumber i
S_{div,s}=-\frac{1}{\epsilon^2}\frac{1}{2\pi}\sqrt{\frac{\lambda
\mu^{2\epsilon}}{(-s)^\epsilon}}
-\frac{1}{\epsilon}\frac{1}{4\pi}(1-\log 2) \sqrt{\frac{\lambda
\mu^{2\epsilon}}{(-s)^\epsilon}} \end{eqnarray}

This answer has the correct general structure (see {\it e.g.} \cite{chapAmp,chapDual}) from field theory expectations. Furthermore, once we use the
strong coupling behavior for the cusp anomalous dimension
\cite{Gubser:2002tv},
$f(\lambda)=\frac{\sqrt{\lambda}}{\pi}+...$ we see that the
leading divergence, as well as the finite piece, have not only
the correct kinematical dependence but also the correct overall
coefficient in order to match the BDS ansatz  \cite{Bern:2005iz,chapDual}. As we will see shortly, this result is actually a consequence of the symmetries of the problem!


\subsection{\texorpdfstring{$T-$}{T-}duality and dual conformal symmetry at strong coupling}

An important ingredient of the previous computation was the
existence of a dual $SO(2,4)$ symmetry \footnote{Actually, this
symmetry was first noticed in perturbative computations
\cite{Drummond:2006rz} and then independently in the strong
coupling computation described here.}, associated to the
isometry group of the dual $AdS_5$ space. This symmetry allowed
the construction of new solutions and fixed somehow the finite
piece of the scattering amplitude. \footnote{Naively, this
conformal symmetry would imply that the amplitude is
independent of
 $s$ and $t$, since they can be sent to arbitrary values by a dual conformal transformation.
 The whole dependence on $s$ and $t$ arises due to the necessity of introducing an
 IR regulator. However after keeping track of the dependence on the
 IR regulator, the amplitude is still determined by the dual conformal symmetry. Hence, this regulator breaks the dual conformal symmetry, but in a controlled
 way!}

To a symmetry we associate a Ward identity and in particular dual conformal symmetry will impose some constraints on the amplitudes. Quite remarkably, this duality was also (actually before!) observed at week coupling and is by now believed to be a duality of scattering amplitudes at all values of the coupling. You can see \cite{chapDual} for a detailed account of this symmetry and the constraints it imposes on the amplitudes.\footnote{These constraints have also been derived at strong coupling \cite{wardstrong,wardstrong2}.} Here we will just mention that dual conformal symmetry fixes the answer for the four-point function to have the form (\ref{finalfour}), actually, to all values of the coupling! and hence its agreement with the BDS ansatz. Furthermore, dual conformal symmetry does not fix the answer for the scattering of more than six gluons, hence, in general, the answer deviates from the BDS ansatz. The need for such a deviation, usually called remainder function, was established in \cite{Alday:2007he,Bern:2008ap}. See \cite{chapDual} for more details.

In the last section we have seen that existence of a dual $AdS$ space, is related to the fact that $AdS_5$ goes to itself after a sequence of four T-dualities, followed by the inversion of the radial coordinate, see (\ref{origads}) vs (\ref{dualads}). This set of T-dualities, however, does not leave the full $AdS_5 \times S^5$ sigma model invariant.  For instance, Buscher rules for T-dualities \cite{Buscher:1987qj} imply a shift on the dilaton of the form
\begin{equation}
\label{bosshift}
\phi \rightarrow \phi +4 \times \log z
\end{equation}
where $z$ is the radial coordinate of the original metric (\ref{origads}). The factor of $4$ is due to the fact that we are making four $T-$dualities. In addition to the usual, "bosonic", $T-$dualities, one can introduce a fermionic $T-$duality \cite{Berkovits:2008ic,Beisert:2008iq}. This duality is a non local redefinition of the fermionic world-sheet fields, very much like the bosonic T-duality is a redefinition of the bosonic fields. These T-dualities change the fields of the sigma model according to precise rules. For instance, each fermionic $T-$duality shifts the dilaton by an amount 
\begin{equation}
\phi \rightarrow \phi - \frac{1}{2}\times \log z
\end{equation}
We see that by doing eight fermionic $T-$dualities we can undo the shift (\ref{bosshift}) on the dilaton. Actually, one can check that  a combination of the four bosonic $T-$dualities plus eight fermionic $T-$dualities maps the full sigma model to itself! Note also that this argument does not depend on the value of the coupling. One of the implications is that the dual model has the same conformal symmetry group as the original, helping to understand the origin of dual conformal symmetry. Actually, as the construction suggests, dual conformal symmetry extends to a full dual super conformal symmetry. In addition, one has a map between the full set of conserved charges of the two models, in such a way that some of the local charge of one model are mapped to non local charges of the dual model, and viceversa, see for instance \cite{Berkovits:2008ic,Beisert:2008iq}.

The structure of dual super conformal symmetry was also seen at weak coupling and is explained in detail in  \cite{chapDual}, for which we refer the reader for more details.


\section{Minimal surfaces on \texorpdfstring{$AdS$}{AdS}}

In the previous section we have seem how the problem of
computing gluon scattering amplitudes at strong coupling
reduces to the computation of the area of certain minimal
surfaces in $AdS$. In this section we show how the
integrability of the system can be used in order to give a
solution to the problem, in the form of a set of integral
equations. We will follow closely
\cite{Alday:2009ga,Alday:2009yn,Alday:2009dv,Alday:2010vh}, see also \cite{Burrington:2009bh,hiss}, to
which we refer the readers for the details. \footnote{Some of the key ideas used below may be found in relation to the study of wall crossing \cite{gmn,gmn2}. Actually, the method of the first paper in \cite{gmn,gmn2} where instrumental in deriving the expression for the eight gluon amplitude at strong coupling in \cite{Alday:2009ga}.} For this review, we
will focus mostly on a particular kinematic configuration, in which
the minimal surfaces are actually embedded into an $AdS_3$
subspace of the full $AdS_5$. However, the full problem has
been solved and it will be briefly mentioned at the end of the section.

The mathematical problem is to find the area of the minimal
surface ending on the boundary of $AdS$ at a given polygon of
light-like edges. The polygon is parametrized by the location of its cusps $x_i$, which
 are null separated, namely $x_{i,i+1}^2=0$.

\begin{figure}[h]
\centering
\includegraphics[scale=0.3]{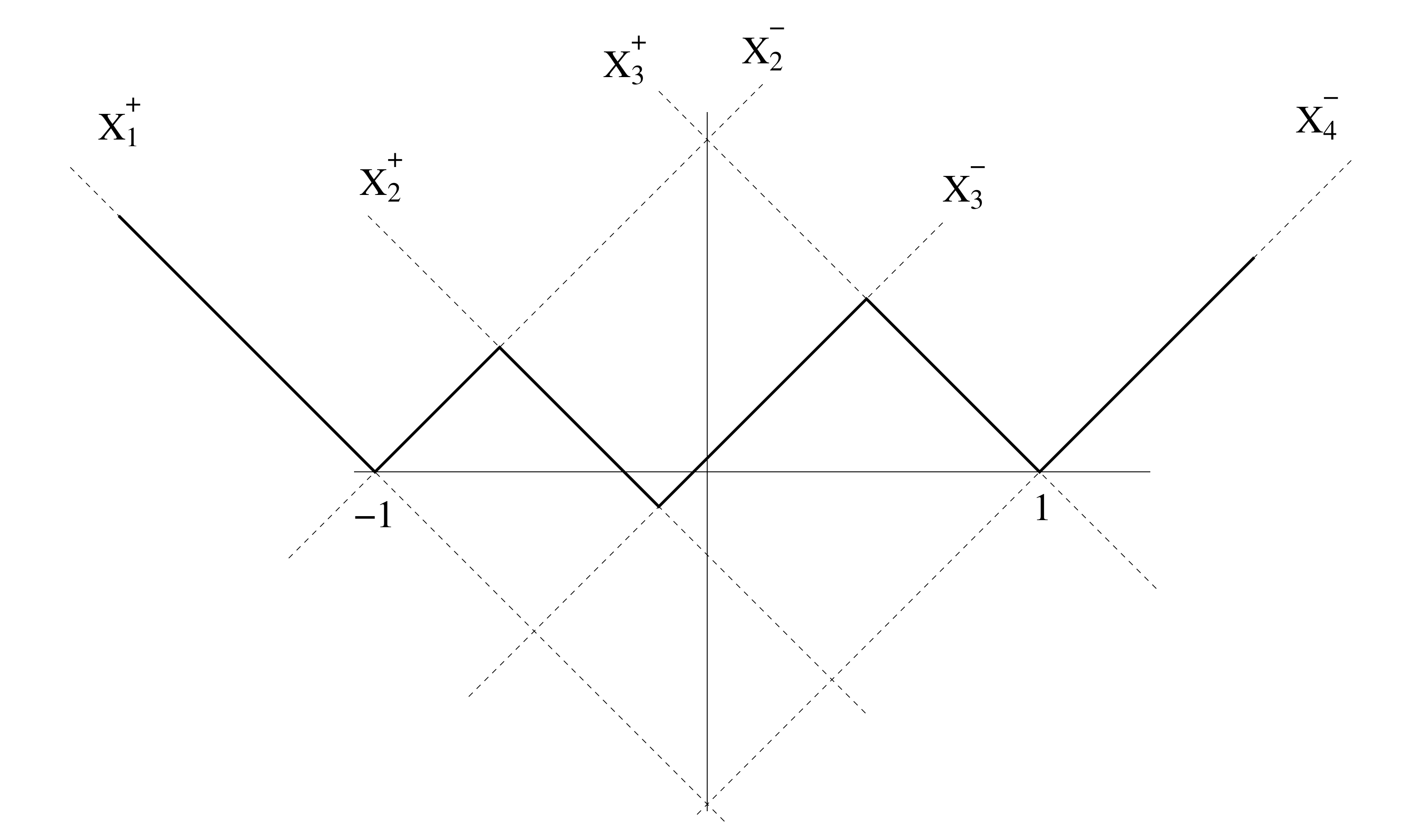}
\caption{A zig-zaged null polygon in $1+1$ dimensions is parametrized by $n$ $x^+_i$ coordinates and $n$ $x^-_i$ coordinates. If you want a closed polygon, you can fold the figure in a cylinder.} \label{octagon}
\end{figure}

We will focus on certain regularized area that is invariant
under conformal transformations. As such, it will depend only
on cross-ratios, of the form $\chi_{ijkl}=\frac{x_{ij}^2
x_{kl}^2}{x_{ik}^2 x_{jl}^2}$. Given the cross-ratios, we want
to compute the area as a function of those. The full problem involves minimal surfaces on
$AdS_5$, in which case there are $3N-15$ independent cross-ratios, where
$N$ is the number of cusps/gluons. We
will restrict to special kinematical configurations in which
the minimal surfaces involved are embedded in $AdS_3$. In this
case, we have $N-6$ independent cross-ratios \footnote{For the general
scattering in four dimensions we have $4N$ coordinates, minus
$N$, since the distance between consecutive points has to be
light-like, minus 15, that is the dimension of the conformal
group $SO(2,4)$. In the case of $AdS_3$, we have $2N-N$ minus
$6$, which is the dimension of $SO(2,2)$.} and the polygon is a
zig-zaged polygon living in one plus one dimensions, which correspond to the boundary of $AdS_3$, see figure
\ref{octagon}.

Since we want a closed contour, and we are in $1+1$ dimensions, we can consider only polygons
with an even number of sides, hence $N=2n$. As one can see in
figure \ref{octagon}, the contour is parametrized by $n$
coordinates $x_i^+$ and  $n$ coordinates $x_i^-$. With each set
of coordinates we can form $n-3$ invariant cross-ratios, of the
form $\chi_{ijkl}^\pm=\frac{x_{ij}^\pm x_{kl}^\pm}{x_{ik}^\pm
x_{jl}^\pm}$.

In order to consider minimal surfaces in $AdS_3$ we need to consider the world-sheet of classical strings on $AdS_3$. This is the subject of the following subsection.

\subsection{Strings on \texorpdfstring{$AdS_3$}{AdS3}}

Classical strings on $AdS_3$ can be described in terms of
embedding coordinates, where $AdS_3$ is the following surface
embedded in $R^{2,2}$
\begin{equation}
\label{Ynorm}
Y.Y \equiv -Y_{-1}^2-Y_0^2+Y_1^2+Y_2^2=-1
\end{equation}
we take the world-sheet to be the whole complex plane.  Since we are interested in classical
solutions, the fields have to satisfy the conformal gauge
equations of motion and the Virasoro constraints
\begin{equation}
\label{Yeom}
\partial \bar \partial Y-(\partial Y . \bar \partial Y) Y=0,~~~~\partial Y . \partial Y= \bar \partial Y . \bar \partial Y=0
\end{equation}
where $\partial Y=\partial_z Y$, etc. An efficient way to focus only in the physical degrees of
freedom, similar to fixing light-cone gauge, is by performing
the so-called Pohlmeyer kind of reduction, see for instance
\cite{po1,po2,po3}, and consider the "reduced" fields
\begin{equation}
\alpha = \log \partial Y . \bar \partial Y,~~~~~p^2 = \partial^2 Y .\partial^2 Y
\end{equation}
As a consequence of the equations of motion and Virasoro
constraints, $p$ can be seen to be a holomorphic function,
$p=p(z)$ while $\alpha(z,\bar z)$ can be seen to satisfy a
generalized version of the Sinh-Gordon equation
\begin{equation}
\label{gsg}
\partial \bar \partial \alpha-e^\alpha+p(z) \bar p(\bar z) e^{-\alpha}=0
\end{equation}
From the definition of the reduced fields, it is clear that they are
invariant under space-time conformal transformation. This means that they describe only the essential part of the problem, without redundancies.

Before proceeding, let us make the following remark. Since
$p(z)$ is a holomorphic function, it is possible to make a
change of coordinates from the $z-plane$ to the  $w-plane$,
where $dw=\sqrt{p(z)} dz$.  In the $w-$plane, after a simple
field redefinition, the generalized Sinh-Gordon equation takes
the usual form
\begin{equation}
\alpha=\hat \alpha+\frac{1}{2} \log p \bar p \rightarrow \partial_w \bar \partial_w \hat \alpha = 2\sinh \hat \alpha
\end{equation}
It would seem that we got rid of all the information on $p(z)$.
However, this is not the case, since the $w-$plane will have in
general a complicated structure (for instance, it will have a
branch cuts, etc, depending on $p(z)$). So, we can choose
between a complicated equation on the complex plane, or a
simple equation on a more complicated space. Depending which
questions we want to answer, one description may be more
convenient than another.  Finally, we are interested in the area of the classical World-sheet. Written in terms of the reduced fields it becomes
\begin{equation}
\label{areaw}
{\cal A}=\int e^\alpha d^2 z=\int e^{\hat \alpha} d^2 w
\end{equation}

\subsection{Classical solutions corresponding to minimal surfaces ending on null polygons}

What are the properties of the holomorphic function $p(z)$ and
$\alpha(z,\bar z)$ for solutions corresponding to minimal surfaces ending on null polygons?
 In order to answer this question we can start by
considering the four cusps solution found in the previous
section and perform the Pohlmeyer reduction. We find
\begin{equation}
p(z)=1,~~~~~\alpha =\hat \alpha=0
\end{equation}
Hence, the four cusps solution simply correspond to the vacuum
solution of the Sinh-Gordon equation!  What about solutions
with a higher number of cusps? First of all we propose that the
field $\alpha$ is regular everywhere, since we are looking for
smooth space-like solutions. Second, we expect a general
solution to be similar to the four cusps solution when
approaching the boundary, so we expect that $\hat \alpha
\rightarrow 0$ as $|z|$ becomes large.

Finally, if we are interested on a minimal surface  ending on a
polygon with $2n$ cusps, we propose $p(z)$ to be a polynomial
of degree $n-2$ \footnote{The motivation for this proposal, comes from the fact that an homogeneous polynomial of degree $n-2$ possesses all the symmetries to correspond to a symmetric polygon of $n$ edges.}
\begin{equation}
p(z)=z^{n-2}+c_{n-4}z^{n-4}+...+ c_1 z+ c_0
\end{equation}
we have used rescalings and translations in order to set the
coefficients of $z^{n-2}$ and $z^{n-3}$ to one and zero
respectively. Such polynomial contains $n-3$ complex
coefficients, or $2n-6$ real coefficients, which exactly agrees
with the amount of expected independent cross ratios for a
polygon with $2n$ cusps!

Summarizing: minimal surfaces ending on a light-like polygon
with $2n$ cusps correspond to a holomorphic polynomial of
degree $n-2$ and a field $\hat \alpha$ satisfying the
Sinh-Gordon equations and with boundary conditions such that it
decays at infinity and diverges logarithmically at the zeroes
of $p(z)$, which amounts to say that $\alpha$ is regular everywhere.

Since $\hat \alpha$ decays at infinity, the integral defining
the area (\ref{areaw}) diverges. We define a regularized area by
subtracting the behavior at infinity
\begin{equation}
A_{reg}=\int (e^{\hat \alpha}-1)d^2w
\end{equation}
 As the reduced fields
are invariant under space-time conformal transformations, the
regularized area will be a function of the cross-ratios only.
\footnote{The full answer would include also the integral of
the one we have subtracted. In order to compute it one would
need to introduce a physical regulator and this part of the
answer will not be conformal invariant. Anyway, its explicitly
form can be worked out and turns out to be quite universal. In
this review we will focus on the "interesting" part of the
answer $A_{reg}$.} The computation of this regularized area is
the main focus of the remaining of this review.

\subsubsection{Reconstructing the space-time solution and its behavior at infinity}

In the following we would to check that the world-sheet we are considering has the desired form. In particular, we would like to understand the shape, in space-time, of the boundary of our world-sheet. For that, we first review a general procedure to reconstruct the world-sheet from the reduced fields, and then study its boundary.

Given an holomorphic function $p(z)$ and a field
$\alpha$ satisfying (\ref{gsg}) it is possible to reconstruct a
space-time solution satisfying (\ref{Yeom}), and
(\ref{Ynorm}). The procedure amounts to solve two auxiliary
linear problems, which we denote as left and right
\begin{eqnarray}
\label{linear}
(d+B^L) \psi^L_{a}=0,~~~~(d+B^R) \psi^R_{\dot a}=0
\end{eqnarray}
where the flat connections $B^{L,R}$ are two by two matrices
constructed from $p(z)$ and $\alpha(z,\bar z)$. For instance
\begin{equation}
B_z^L=\left(\begin{matrix} \partial \alpha/4 & \frac{1}{\sqrt{2}}e^{\alpha/2} \\ \frac{1}{\sqrt{2}}p e^{-\alpha/2} & -\partial \alpha/4 \end{matrix}  \right)
\end{equation}
we denote different components of the connections by
$B^L_{\alpha \beta}$ and $B^R_{\dot \alpha \dot \beta}$. On the
other hand, the indices $a$ and $\dot a$ in (\ref{linear})
denote independent solutions of the auxiliary linear problems.
Each $\psi_a^L$ or $\psi_{\dot a}^R$ is then a doublet. We
denote the components of this doublet by $\psi^L_{\alpha,a}$,
etc.

Given the solutions of these two auxiliary linear problems, one
can show that the space-time coordinates are simply given by
\begin{equation}
Y_{a,\dot a}=\left(\begin{matrix} Y_{-1}+Y_2 & Y_1-Y_0 \\ Y_1+Y_0 & Y_{-1}+Y_2 \end{matrix}  \right)_{a,\dot a}=\psi^L_{\alpha,a} \delta^{\alpha \dot \beta} \psi^R_{\dot \beta, \dot a}
\end{equation}
 One can show that $Y$
constructed this way satisfies all the required properties. If we see $\psi^L$ and $\psi^R$ as two by two matrices, then
the space-time coordinates would be given by $Y = (\psi^L)^T
\psi^R$. On the other hand, note that given a solution to the
left problem, $\psi^L$, then $\psi^L U^T$ is an equally good
solution, and the same happens with the right problem. Hence,
given $Y$, we obtain a family of space-time solutions $U Y V$.
These are nothing but the space-time conformal transformations.

Now we would like to understand the behavior of the solutions of
the linear auxiliary problems for very large values of $|z|$,
or $|w|$. This will tell us the behavior of the world-sheet
near its boundary. Let us start, by simplicity, with the case
of a homogeneous polynomial, $p(z)=z^{n-2}$. Hence $w \approx
z^{n/2}$. As a result, as we go once around the $z-$plane, we
go around the $w-$plane $n/2$ times.

\begin{figure}[h]
\centering
\includegraphics[scale=0.3]{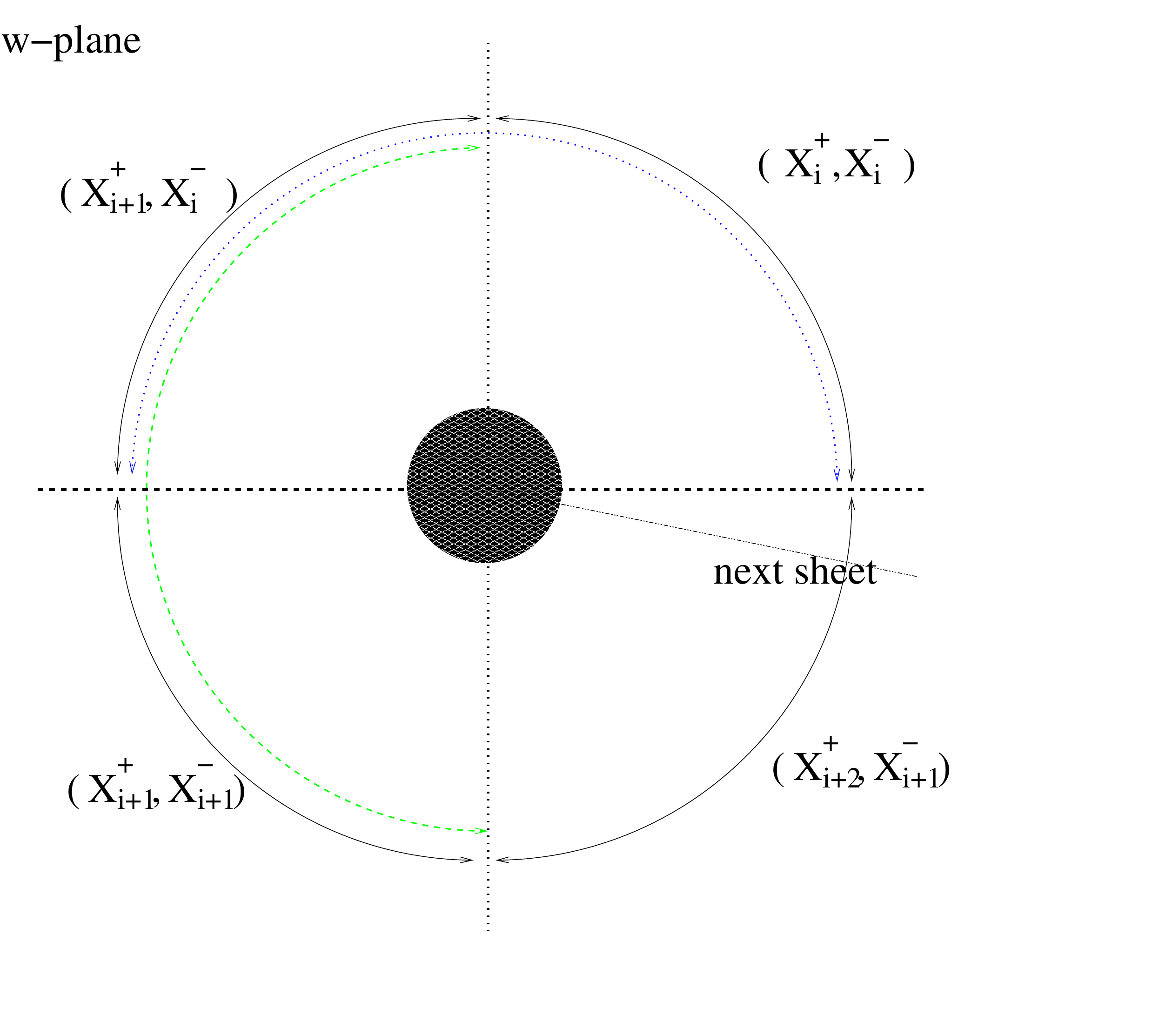}
\caption{When looking at the left problem, (each sheet of) the $w-$plane is naturally divided into two parts, according to the sign of $Re(w)$. In the same way, when looking at the right problem, the $w-$plane is naturally divided into two parts, according to the sign of $Im(w)$. Hence, the $w-$plane is naturally divided into four quadrants. Large values of $|w|$ in each of these angular sectors correspond to a cusp.} \label{wplane}
\end{figure}

Due to the boundary conditions for the reduced fields, the flat connections $B^{L,R}$
drastically simplify at infinity and we can solve the auxiliary
linear problems. A general solution will be of the form
\begin{eqnarray}
\label{approxsol}
\psi^L_a \approx c_a^+ \left( \begin{matrix} 1 \\0  \end{matrix} \right) e^{w +\bar w}+c_a^- \left( \begin{matrix} 0 \\1  \end{matrix} \right)e^{-(w+\bar w)} \cr
\psi^R_{\dot a} \approx d_{\dot a}^+ \left( \begin{matrix} 1 \\0  \end{matrix} \right) e^{\frac{w -\bar w}{i}}+d_{\dot a}^- \left( \begin{matrix} 0 \\1  \end{matrix} \right)e^{-\frac{w -\bar w}{i}}
\end{eqnarray}
The $w$-plane is naturally divided into quadrants, see figure
\ref{wplane}.
In each quadrant one of the two solutions of each problem (left
and right) dominates. For instance, in the upper right
quadrant, the solution proportional to $c_a^+$ dominates in the
left problem, while the solution proportional to $d_{\dot a}^+$
dominates in the right problem. This means that for large
values of $|w|$, the whole quadrant corresponds to a single
point in the boundary, given by $Y_{a \dot a} \approx (Large)
\times c_a^+ d_{\dot a}^+$.

As we change quadrant, one and only one of the two dominant
solutions change and we jump a light-like distance to the next
cusp. In each quadrant/cusp we can write
\begin{equation}
Y_{a,\dot {a}} \approx \lambda_a \tilde \lambda_{\dot a}
\end{equation}
where $\lambda$ is given by the leading contribution to the
left problem and $\tilde \lambda$ by the leading contribution
to the right problem. As we change quadrant, one of the two
solutions, $\lambda$ or $\tilde \lambda$, changes. As we go
around the $w-$plane $n/2$ times, we get the expected $2n$
cusps!

In the general case in which the polynomial is not homogeneous,
the picture is very much the same. In general, the degree of
the polynomial determines the number of cusps, while the
coefficients on the polynomial determine the shape of the
polygon.

Let us finish this section with a very important observation. Since we know that the classical equations describing strings on $AdS_5 \times S^5$ are integrable, see for instance \cite{Bena:2003wd},  we expect the present problem to be integrable as well. Indeed, it is possible to
introduce a spectral parameter $\zeta$ 
\begin{eqnarray}
B_z \rightarrow B_z(\zeta)=\frac{1}{4}\left(\begin{matrix} \partial \alpha &0 \\0& -\partial \alpha \end{matrix} \right) + \frac{1}{\zeta} \frac{1}{\sqrt{2}}\left(\begin{matrix} 0&e^{\alpha/2} \\ p e^{-\alpha/2}& 0\end{matrix} \right) \\
B_{\bar z} \rightarrow B_{\bar z}(\zeta) =\frac{1}{4} \left(\begin{matrix}- \bar \partial \alpha &0 \\0& \bar \partial \alpha \end{matrix} \right) + \zeta \frac{1}{\sqrt{2}}\left(\begin{matrix} 0&\bar p e^{- \alpha/2} \\  e^{\alpha/2}& 0\end{matrix} \right)
\end{eqnarray}
such that the flat connections are still flat, namely the satisfy $\partial B_{\bar z}-\bar \partial B_z+[B_z,B_{\bar z}]=0$, for all values of $\zeta$. The introduction of the spectral parameter, also allows to
study both linear problems in a unified manner. The left and
right connections are just particular cases of the above flat
connection, more precisely
\begin{equation}
B(\zeta =1) = B^L,~~~~B(\zeta=i)=B^R
\end{equation}
The existence of the spectral parameter played a key role in the spectrum problem, see for instance \cite{SchaferNameki:2010jy}. In that context, the key object are the eigenvalues of the monodromy matrix constructed out of the flat connection. In the present case the key object are certain cross-ratios constructed from the holonomy of the connections.

\subsection{Y-system for minimal surfaces}

Let us focus on the left problem. We see that each sheet on the
$w-$plane is naturally divided into two sectors, one with
$Re(w)>0$ and the other with $Re(w)<0$. In each sector the
small solution is well defined (up to a normalization
constant). On the other hand, the large solution is not, as we
can add to it a part of the small solution. Let us then
introduce the following terminology:
\begin{itemize}
\item The $w-$plane is divided into $n$ sectors, since each sheet contains two sectors. We label this sectors by $i=0,...,n-1$.
\item We call $s_i^L$ the small solution at the $i-th$
    sector. This is the solution with the
    fastest decay along the line in the center of the
    $i-th$ sector, for increasing $|w|$.
\end{itemize}
In order to understand why these small solutions are important, we need to introduce a new element. Given that our connections are $SL(2)$ matrices, we can introduce a $SL(2)$ invariant product
\begin{equation}
\label{psinor}
\psi_a^L \wedge \psi_b^L \equiv \epsilon^{\alpha \beta} \psi^L_{\alpha,a}  \psi^L_{\beta,b} = \epsilon_{ab}
\end{equation}
The second equality corresponds to setting a normalization factor for our solutions. This can be done since one can check the above product is independent on the world-sheet coordinate $z$. 

As already seen, the location of the cusps is determined by the
large solutions. The large component of a solution, on a given
sector, can be extracted by using the small solution on such
sector and the $SL(2)$ invariant product just introduced, more
precisely
\begin{equation}
\psi_a^L \wedge s_i^L \approx \lambda_a^i
\end{equation}
where the symbol $\approx$ means up to factors that will cancel out in the final expression for the cross-ratios. How do we construct space-time cross-ratios? we have see that
the location of the cusps is given by $Y^i_{a \dot
a}=\lambda_a^i \tilde \lambda_{\dot a}^i$. The space time
cross-ratios involve distances like
\begin{equation}
Y^i . Y^j = \epsilon^{ab} \epsilon^{\dot a \dot b} Y^i_{a \dot a} Y^j_{b \dot b}=<\lambda^i \lambda^j> <\tilde \lambda^i \tilde \lambda^j>,~~~~~<\lambda^i \lambda^j> =\epsilon^{ab} \lambda^i_a \lambda^j_b
\end{equation}
Given the normalization condition (\ref{psinor}), one can
easily show
\begin{equation}
<\lambda^i \lambda^j> \approx < (\psi_a^L \wedge s_i^L)(\psi_a^L \wedge s_j^L)> = s_i^L \wedge s_j^L
\end{equation}
Which means that space-time cross-ratios can be constructed from inner products of the small solutions in the corresponding sectors! more precisely
\begin{equation}
\frac{x_{ij}^+ x_{kl}^+ }{x_{ik}^+ x_{jl}^+}=\frac{(s_i^L \wedge s_j^L)(s_k^L \wedge s_l^L)}{(s_i^L \wedge s_k^L)(s_j^L \wedge s_l^L)}
\end{equation}
Small solutions are defined up to a normalization constant.
Note that such normalization constants cancel out when
computing cross-ratios.

The strategy we will follow is to introduce the spectral
parameter $\zeta$ as shown in the previous section and study
the small solutions of the corresponding connection
\begin{equation}
(d+B(\zeta))s_i(\zeta)=0
\end{equation}
then, we can consider the cross-ratios as a function of such spectral parameter
\begin{equation}
\chi_{ijkl}(\zeta)=\frac{(s_i \wedge s_j)(s_k \wedge s_l )}{(s_i \wedge s_k )(s_j \wedge s_l)}
\end{equation}
The physical cross-ratios are then obtained by setting the
spectral parameter to appropriate values
\begin{equation}
\chi_{ijkl}(\zeta=1)=\chi_{ijkl}^+,~~~~~\chi_{ijkl}(\zeta=i)=\chi_{ijkl}^-
\end{equation}
A very important property of the flat connection $B(\zeta)$ is
that it possesses a $Z_2$ symmetry: $B(e^{i\pi} \zeta)=\sigma_3 B(\zeta) \sigma_3$, where $\sigma_3$ is the usual Pauli matrix. This symmetry
allows to relate small solutions at different values of the
spectral parameter, for instance $s_{i+1}(\zeta)=\sigma_3
s_i(e^{i\pi} \zeta)$, and in particular, it implies
\begin{equation}
\label{z2rel}
s_i \wedge s_j (e^{i \pi} \zeta)=s_{i+1} \wedge s_{j+1}(\zeta)
\end{equation}
This identity is crucial in deriving the equations below.
Besides, in order to simplify subsequent expressions, we will
assume $s_i \wedge s_{i+1}=1$.

Now we have all the elements to derive the so called Hirota
equations and the Y-system equations. The trick is to choose
$s_0$ and $s_1$ as a complete basis of flat sections, and
express two arbitrary consecutive small solutions $s_k$ and
$s_{k+1}$ in terms of these
\begin{eqnarray}
s_k& =& (s_k \wedge s_1) s_0-(s_k \wedge s_0) s_1 \\
s_{k+1} &=& (s_{k+1} \wedge s_1) s_0-(s_{k+1} \wedge s_0) s_1
\end{eqnarray}
Next, use (\ref{z2rel}) in order to express every wedge as a wedge involving $s_0$ and consider $1=s_k \wedge s_{k+1}$, we obtain
\begin{equation}
-(s_{k-1} \wedge s_0)^{++} (s_{k+1} \wedge s_0)+(s_k \wedge s_0)^{++} (s_k \wedge s_0)=1
\end{equation}
where we have introduced the notation $f^\pm=f(e^{\pm i
\pi/2}\zeta)$, $f^{++}=f(e^{ i \pi}\zeta)$, etc. Let us
introduce $T_k=s_0 \wedge s_{k+1} (e^{-i (k+1) \pi/2} \zeta)$.
In terms of these we obtain
\begin{equation}
T_s^+ T_s^-=T_{s+1}T_{s-1} +1
\end{equation}
which has the form of the so called Hirota equations! from the
definition of $T_s$, we see that it is non trivial for
$s=0,...,n-2$. The $Y-$system equations can be obtained by
introducing $Y_s \equiv T_{s-1} T_{s+1}$
\begin{equation}
\label{Ysystem}
Y_s^+ Y_s^-=(1+Y_{s+1})(1+Y_{s-1})
\end{equation}
$Y_s$ is non trivial for $s=1,...,n-3$. Note that this agrees
with the amount of (complex) cross-ratios of our scattering
problem. These are functional equations for $Y_s(\zeta)$ and
are valid for any value of $\zeta$. Note that they followed from
a chain of rather trivial facts!

One could reintroduce the normalized factors $s_i \wedge
s_{i+1}$ and check that the $Y-$functions are given by the
usual cross-ratios introduced above. The physical cross-ratios,
are then obtained by evaluating $Y_s(\zeta)$ at $\zeta=1$ and
$\zeta=i$.

Such equations are not the whole story. In particular, note
that that the holomorphic function $p(z)$ does not enter at all
in such equations! The point is the following. There are many
solutions to such equations. The correct solution is then
picked by specifying the analytic properties and boundary
conditions of $Y_s(\zeta)$ as we move on the $\zeta-$plane. 
This is how the information about the holomorphic polynomial
enters and will be the subject of the following section.

\subsection{Integral equations}

In order to pick the appropriate solution to the $Y-$system
equations (\ref{Ysystem}) we need to specify the analytic
properties of $Y_s(\zeta)$. By analyzing the auxiliary linear
problems and the definition of $Y_s(\zeta)$ one can show that
$Y_s(\zeta)$ are analytic away from $\zeta=0,\infty$. On the
other hand, As $\zeta \rightarrow 0,\infty$, the flat
connection simplifies and the inverse problem can be solved by
using a $WKB$ approximation, where the role of $\hbar$ is
played by $\zeta$ or $1/\zeta$ . By calling $\zeta=e^\theta$,
one can show that for large $\theta$ the solution behaves as \footnote{Note that even though we used the WKB approximation, this is the behavior of the exact solution.}
\begin{equation}
\log Y_s \approx - m_s \cosh \theta+...
\end{equation}
where $m_a$ is given by the periods of  $p(z)^{1/2}$ along the
cycles $\gamma_a$, namely $m_a \approx - \oint_{\gamma_a}
\sqrt{p(z)} dz$. This is how the information of the polynomial
$p(z)$ enters into the problem. This periods are usually
complex, and there are $n-3$ of them, which exactly agrees with
the quantity of expected cross-ratios. These $m_a$ should be
seen as the boundary conditions for the above equations.

The strategy now is well known from the study of integrable
systems. We can combine the $Y-$system equations with the
analytic properties and boundary conditions for the
$Y-$functions, in order to write a system of integral equations
for them. The solutions to these integral equations will
automatically satisfy the $Y-$system equations and have the
required boundary conditions. The system of integral equations
is given by \footnote{Considering $l_s \equiv \log(Y_s/e^{-m_s \cosh \theta})$, which is analytic in the strip $Im(\theta)<\pi/2$ and vanishes as $\theta$ approaches infinite. The integral equations can be obtained by convoluting the equation $l_s^++l_s^-=\log(1+Y_{s+1})(1+Y_{s-1})$ with the kernel in (\ref{tba}).}
\begin{equation}
\label{tba}
\log Y_s= -m_s \cosh \theta +\frac{1}{2\pi}  \int_{-\infty}^\infty \frac{1}{\cosh(\theta-\theta')} \log (1+Y_{s+1}(\theta'))(1+Y_{s-1}(\theta'))
\end{equation}
%
%
%
The system of equations (\ref{tba}) has the form of TBA
equations, that arise when studying integrable models in finite
volume, see {\it e.g.} \cite{Zamolodchikov:1989cf}. Even
though, for the sake of clarity, some overall coefficients have
been suppressed in the derivation of these equations, the final
form of the equations is given with all the correct
coefficients. From the TBA point of view, the parameters $m_a$
enter as masses. Once the masses are given, the solution of the
above system is unique. The physical cross-ratios can the been
read off by looking at $Y_s(\theta)$ for appropriate values of
$\theta$.

These integral equations can also be written in terms of the physical cross-ratios $y_s^{+}=Y_s(\zeta=1)$ and $y_s^{-}=Y_s(\zeta=i)$ only, without resorting to the auxiliary parameters $m_s$. In order to achieve that, one simple evaluates (\ref{tba}) at the physical values of the spectral parameter in order to eliminate the masses, see \cite{Alday:2010ku}. We obtain
\begin{equation}
\label{tbanew}
\log Y_s= \cosh \theta \log y_s^+ -i \sinh \theta \log y_s^- +\int d\theta' \frac{\sinh 2\theta}{\cosh(\theta'-\theta)\sinh 2\theta'}  \log (1+Y_{s+1}(\theta'))(1+Y_{s-1}(\theta'))
\end{equation}
Note that having solved (\ref{tbanew}), we could read off the masses from the asymptotic behavior of the solutions.

How do we compute the regularized
area, once we have solved the above system of integral
equations? It turns out that the area can be written in terms
of the $Y-$functions in a very simple form
\begin{equation}
\label{regarea}
A_{reg}=\sum_s \int d\theta \frac{m_s}{2\pi} \cosh \theta \log(1+Y_s(\theta))
\end{equation}
In order to derive this expression, one expands the $Y$ functions a few orders for small and for large values of $\zeta$. The expansion coefficients are written in terms of period integrals that also appear in the expression for the area, see \cite{Alday:2010vh} for details. This expression has exactly the form of the free energy of the TBA system. \footnote{It can be shown \cite{Alday:2010ku} that actually this area coincides with the extremum of the Yang-Yang functional for the modified TBA equations (\ref{tbanew})}.

The strategy to solve the full problem is then clear. For a choice of the cross-ratios we solve the integral equations (\ref{tbanew}), and from their solution we compute the area (\ref{regarea}). Hence, we have
the area for these values of the cross-ratios.

In this review we have treated in detail the case of minimal surfaces in $AdS_3$. However, the general case of minimal surfaces in $AdS_5$ can also be solved \cite{Alday:2010vh}. Much of what we have said can be carried out for the general case. In this case we get a bigger system of $Y-$functions, denoted by $Y_{a,s}$, where $a=1,23$ and $s=1,...,N-5$. Note that their number equals the number of independent cross-ratios. Very much as before, one can obtain $Y-$system equations, which supplemented with the appropriate boundary conditions can be written as a system of integral equations. Again, this system of equations has the form of a TBA system, and the regularized area coincides with the free energy of such system.

These equations can be solved numerically, see for instance \cite{Alday:2010vh}. On the other hand, it is very hard to find analytical solutions. However, some limits, for instance the so called small masses/CFT limit and the large masses limit , are more tractable, see \cite{limits,limits2,limits3}.


\section{Conclusions}

We reviewed the computation of scattering amplitudes of planar
maximally super-symmetric Yang-Mills at strong coupling. By
using the $AdS/CFT$ duality the problem boils down to the
computation of the area of certain minimal surfaces on $AdS$.

Then we showed how the integrability of the model can be efficiently used in
order to give an answer for the problem in terms of a set of
integral equations. Integrability allows to introduce a one parameter deformation (the spectral parameter $\zeta$) and study such deformed problem. One can then write down a system of functional equations, or $Y-$system, valid for any value of $\zeta$. One can combine these functional equations with the knowledge of the analytic behavior of the $Y-$functions in the $\zeta-plane$, in order to write a set of integral equations which can be solved iteratively, and give the desired answer. There are many directions one could try to follow, some of the most interesting are the following

\begin{itemize}
\item It would be nice to find a physical connection between the integrable system that the TBA equations describe and the original integrable system.
\item It would be very interesting to extend the present construction to the full quantum problem. As a first step, one could try to compute one loop (from the strong coupling point of view) corrections to the above picture. This would allow, for instance, to distinguish between different amplitudes.
\item It would also be interesting to look for similar structures (for instance, the analogous of the spectral parameter, etc) in perturbative computations. Related to this, in \cite{Alday:2010ku} and subsequent papers, an operator product expansion for Wilson loops have been developed. This allows to use certain tools of integrability \cite{Basso:2010in} in order to make predictions at all values of the coupling.
\item One could hope that similar technology can be applied to related problems. One such problem is the computation of form factors, in which progress have been made recently, see \cite{Alday:2007he}\cite{Maldacena:2010kp}\cite{Brandhuber:2010ad}.
\end{itemize}

{\bf Acknowledgments}

We would like to thank Davide Gaiotto, Amit Sever, Pedro Vieira
and specially Juan Maldacena for collaboration on the material
exposed in this review.

\phantomsection
\addcontentsline{toc}{section}{\refname}

\begin{thebibliography}{00}    

\bibitem[V.1]{chapAmp}
R.~Roiban,
``Review of AdS/CFT Integrability, Chapter V.1: Scattering Amplitudes
  -- a Brief Introduction,''
\arxivref{1012.4001}{arXiv:1012.4001 [hep-th]}.


\bibitem[V.2]{chapDual}
J.~Drummond,
``Review of AdS/CFT Integrability, Chapter V.2: Dual Superconformal
  Symmetry,''
\arxivref{1012.4002}{arXiv:1012.4002 [hep-th]}.


\bibitem{Maldacena:1997re}
  J.~M.~Maldacena,
  ``The large N limit of superconformal field theories and supergravity,''
  Adv.\ Theor.\ Math.\ Phys.\  {\bf 2} (1998) 231
  [Int.\ J.\ Theor.\ Phys.\  {\bf 38} (1999) 1113]
  [arXiv:hep-th/9711200].

\bibitem{Rey:1998ik}
  S.~J.~Rey and J.~T.~Yee,
  ``Macroscopic strings as heavy quarks in large N gauge theory and  anti-de
  Sitter supergravity,''
  Eur.\ Phys.\ J.\  C {\bf 22} (2001) 379
  [arXiv:hep-th/9803001].

\bibitem{Maldacena:1998im}
  J.~M.~Maldacena,
  ``Wilson loops in large N field theories,''
  Phys.\ Rev.\ Lett.\  {\bf 80} (1998) 4859
  [arXiv:hep-th/9803002].

\bibitem{Alday:2007hr}
  L.~F.~Alday and J.~M.~Maldacena,
  ``Gluon scattering amplitudes at strong coupling,''
  JHEP {\bf 0706} (2007) 064
  [arXiv:0705.0303 [hep-th]].


\bibitem{Alday:2009zm}
  L.~F.~Alday, J.~M.~Henn, J.~Plefka and T.~Schuster,
  ``Scattering into the fifth dimension of N=4 super Yang-Mills,''
  JHEP {\bf 1001} (2010) 077
  [arXiv:0908.0684 [hep-th]].



\bibitem{Gross:1987ar}
  D.~J.~Gross and P.~F.~Mende,
  ``String Theory Beyond the Planck Scale,''
  Nucl.\ Phys.\  B {\bf 303} (1988) 407.



\bibitem{McGreevy:2008zy}
  J.~McGreevy and A.~Sever,
  ``Planar scattering amplitudes from Wilson loops,''
  JHEP {\bf 0808} (2008) 078
  [arXiv:0806.0668 [hep-th]].

\bibitem{finitetemp}
  K.~Ito, H.~Nastase and K.~Iwasaki,
  ``Gluon scattering in N = 4 super Yang-Mills at finite temperature,''
  Prog.\ Theor.\ Phys.\  {\bf 120} (2008) 99
  [arXiv:0711.3532 [hep-th]].

\bibitem{finitetemp2}
  G.~Georgiou and D.~Giataganas,
  ``Gluon Scattering Amplitudes in Finite Temperature Gauge/Gravity
  Dualities,''
  arXiv:1011.6339 [hep-th].

\bibitem{Dorn:2009hs}
  H.~Dorn, N.~Drukker, G.~Jorjadze and C.~Kalousios,
  ``Space-like minimal surfaces in AdS x S,''
  JHEP {\bf 1004} (2010) 004
  [arXiv:0912.3829 [hep-th]].

\bibitem{quarks}
  J.~McGreevy and A.~Sever,
  ``Quark scattering amplitudes at strong coupling,''
  JHEP {\bf 0802} (2008) 015
  [arXiv:0710.0393 [hep-th]].
 
\bibitem{quarks2}
Z.~Komargodski and S.~S.~Razamat,
  JHEP {\bf 0801} (2008) 044
  [arXiv:0707.4367 [hep-th]].

\bibitem{quarks3}
  E.~Barnes and D.~Vaman,
  ``Massive quark scattering at strong coupling from AdS/CFT,''
  Phys.\ Rev.\  D {\bf 81} (2010) 126007
  [arXiv:0911.0010 [hep-th]].












\bibitem{Kruczenski:2002fb}
  M.~Kruczenski,
  ``A note on twist two operators in N = 4 SYM and Wilson loops in Minkowski
  signature,''
  JHEP {\bf 0212} (2002) 024
  [arXiv:hep-th/0210115].

\bibitem{Gubser:2002tv}
  S.~S.~Gubser, I.~R.~Klebanov and A.~M.~Polyakov,
  ``A semi-classical limit of the gauge/string correspondence,''
  Nucl.\ Phys.\  B {\bf 636} (2002) 99
  [arXiv:hep-th/0204051].
  
\bibitem{Bern:2005iz}
  Z.~Bern, L.~J.~Dixon and V.~A.~Smirnov,
  ``Iteration of planar amplitudes in maximally supersymmetric Yang-Mills
  theory at three loops and beyond,''
  Phys.\ Rev.\  D {\bf 72}, 085001 (2005)
  [arXiv:hep-th/0505205].
  






  
\bibitem{Drummond:2006rz}
  J.~M.~Drummond, J.~Henn, V.~A.~Smirnov and E.~Sokatchev,
  ``Magic identities for conformal four-point integrals,''
  JHEP {\bf 0701} (2007) 064
  [arXiv:hep-th/0607160].


\bibitem{wardstrong}
  Z.~Komargodski,
  ``On collinear factorization of Wilson loops and MHV amplitudes in N=4 SYM,''
  JHEP {\bf 0805} (2008) 019
  [arXiv:0801.3274 [hep-th]].

\bibitem{wardstrong2}
  L.~F.~Alday,
  ``Lectures on Scattering Amplitudes via AdS/CFT,''
  Fortsch.\ Phys.\  {\bf 56} (2008) 816
  [arXiv:0804.0951 [hep-th]].
  

\bibitem{Alday:2007he}
  L.~F.~Alday and J.~Maldacena,
  ``Comments on gluon scattering amplitudes via AdS/CFT,''
  JHEP {\bf 0711} (2007) 068
  [arXiv:0710.1060 [hep-th]].
  
\bibitem{Bern:2008ap}
  Z.~Bern, L.~J.~Dixon, D.~A.~Kosower, R.~Roiban, M.~Spradlin, C.~Vergu and A.~Volovich,
  ``The Two-Loop Six-Gluon MHV Amplitude in Maximally Supersymmetric Yang-Mills
  Theory,''
  Phys.\ Rev.\  D {\bf 78} (2008) 045007
  [arXiv:0803.1465 [hep-th]].
  
\bibitem{Buscher:1987qj}
  T.~H.~Buscher,
  ``Path Integral Derivation of Quantum Duality in Nonlinear Sigma Models,''
  Phys.\ Lett.\  B {\bf 201}, 466 (1988).
  
\bibitem{Berkovits:2008ic}
  N.~Berkovits and J.~Maldacena,
  ``Fermionic T-Duality, Dual Superconformal Symmetry, and the Amplitude/Wilson
  Loop Connection,''
  JHEP {\bf 0809}, 062 (2008)
  [arXiv:0807.3196 [hep-th]].

\bibitem{Beisert:2008iq}
  N.~Beisert, R.~Ricci, A.~A.~Tseytlin and M.~Wolf,
  ``Dual Superconformal Symmetry from AdS5 x S5 Superstring Integrability,''
  Phys.\ Rev.\  D {\bf 78}, 126004 (2008)
  [arXiv:0807.3228 [hep-th]].

 \bibitem{Alday:2009ga}
  L.~F.~Alday and J.~Maldacena,
  ``Minimal surfaces in AdS and the eight-gluon scattering amplitude at strong
  coupling,''
  arXiv:0903.4707 [hep-th].

\bibitem{Alday:2009yn}
  L.~F.~Alday and J.~Maldacena,
  ``Null polygonal Wilson loops and minimal surfaces in Anti-de-Sitter space,''
  JHEP {\bf 0911}, 082 (2009)
  [arXiv:0904.0663 [hep-th]].


\bibitem{Alday:2009dv}
  L.~F.~Alday, D.~Gaiotto and J.~Maldacena,
  ``Thermodynamic Bubble Ansatz,''
  arXiv:0911.4708 [hep-th].

\bibitem{Alday:2010vh}
  L.~F.~Alday, J.~Maldacena, A.~Sever and P.~Vieira,
  ``Y-system for Scattering Amplitudes,''
  arXiv:1002.2459 [hep-th].

\bibitem{Burrington:2009bh}
  B.~A.~Burrington and P.~Gao,
  ``Minimal surfaces in AdS space and Integrable systems,''
  JHEP {\bf 1004} (2010) 060
  [arXiv:0911.4551 [hep-th]].
  
\bibitem{hiss}
  Y.~Hatsuda, K.~Ito, K.~Sakai, Y.~Satoh,
  ``Thermodynamic Bethe Ansatz Equations for Minimal Surfaces in AdS(3),''
  JHEP {\bf 1004}, 108 (2010).
  [arXiv:1002.2941 [hep-th]].

\bibitem{gmn}
  D.~Gaiotto, G.~W.~Moore and A.~Neitzke,
  ``Four-dimensional wall-crossing via three-dimensional field theory,''
  Commun.\ Math.\ Phys.\  {\bf 299} (2010) 163
  [arXiv:0807.4723 [hep-th]].

\bibitem{gmn2}
  D.~Gaiotto, G.~W.~Moore and A.~Neitzke,
  ``Wall-crossing, Hitchin Systems, and the WKB Approximation,''
  arXiv:0907.3987 [hep-th].

\bibitem{po1}
  H.~J.~De Vega and N.~G.~Sanchez,
  ``Exact Integrability Of Strings In D-Dimensional De Sitter Space-Time,''
  Phys.\ Rev.\  D {\bf 47}, 3394 (1993).

\bibitem{po2}
  A.~Jevicki, K.~Jin, C.~Kalousios and A.~Volovich,
  ``Generating AdS String Solutions,''
  JHEP {\bf 0803}, 032 (2008)
  [arXiv:0712.1193 [hep-th]].

\bibitem{po3}
  H.~Dorn,
  ``Some comments on spacelike minimal surfaces with null polygonal boundaries
  in $AdS_m$,''
  JHEP {\bf 1002}, 013 (2010)
  [arXiv:0910.0934 [hep-th]].
  
\bibitem{Zamolodchikov:1989cf}
  A.~B.~Zamolodchikov,
  ``Thermodynamic Bethe Ansatz In Relativistic Models. Scaling Three State
  Potts And Lee-Yang Models,''
  Nucl.\ Phys.\  B {\bf 342}, 695 (1990).

\bibitem{Bena:2003wd}
  I.~Bena, J.~Polchinski and R.~Roiban,
  ``Hidden symmetries of the AdS(5) x S**5 superstring,''
  Phys.\ Rev.\  D {\bf 69} (2004) 046002
  [arXiv:hep-th/0305116].

\bibitem{SchaferNameki:2010jy}
  S.~Schafer-Nameki,
  ``Review of AdS/CFT Integrability, Chapter II.4: The Spectral Curve,''
  arXiv:1012.3989 [hep-th].




\bibitem{limits}
  Y.~Hatsuda, K.~Ito, K.~Sakai, Y.~Satoh,
  ``Six-point gluon scattering amplitudes from Z4-symmetric integrable model,''
  JHEP {\bf 1009}, 064 (2010).
  [arXiv:1005.4487 [hep-th]].

\bibitem{limits2}
  Y.~Hatsuda, K.~Ito, K.~Sakai, Y.~Satoh,
  ``g-functions and gluon scattering amplitudes at strong coupling,''
  [arXiv:1102.2477 [hep-th]].

\bibitem{limits3}
J.~Bartels, J.~Kotanski, V.~Schomerus,
  ``Excited Hexagon Wilson Loops for Strongly Coupled N=4 SYM,''
  JHEP {\bf 1101}, 096 (2011).
  [arXiv:1009.3938 [hep-th]].





\bibitem{Alday:2010ku}
  L.~F.~Alday, D.~Gaiotto, J.~Maldacena, A.~Sever and P.~Vieira,
  ``An Operator Product Expansion for Polygonal null Wilson Loops,''
  arXiv:1006.2788 [hep-th].

\bibitem{Basso:2010in}
  B.~Basso,
  ``Exciting the GKP string at any coupling,''
  arXiv:1010.5237 [hep-th].
  
\bibitem{Maldacena:2010kp}
  J.~Maldacena and A.~Zhiboedov,
  ``Form factors at strong coupling via a Y-system,''
  JHEP {\bf 1011} (2010) 104
  [arXiv:1009.1139 [hep-th]].
  
\bibitem{Brandhuber:2010ad}
  A.~Brandhuber, B.~Spence, G.~Travaglini and G.~Yang,
  ``Form Factors in N=4 Super Yang-Mills and Periodic Wilson Loops,''
  JHEP {\bf 1101} (2011) 134
  [arXiv:1011.1899 [hep-th]].

\end{thebibliography}

\end{document}